\documentclass[fleqn,usenatbib]{mnras}

\usepackage[T1]{fontenc}

\DeclareRobustCommand{\VAN}[3]{#2}
\let\VANthebibliography\thebibliography
\def\thebibliography{\DeclareRobustCommand{\VAN}[3]{##3}\VANthebibliography}

\newcommand{\lya}{Ly$\alpha\,$}

\newcommand{\cii}{[\ion{C}{ii}]}
\newcommand{\civ}{\ion{C}{iv}}
\newcommand{\mgii}{\ion{Mg}{ii}}
\newcommand{\feii}{\ion{Fe}{ii}}

\newcommand{\nv}{\ion{N}{v}}

\newcommand{\hb}{H$\beta\,$}

\usepackage{graphicx}	
\usepackage{amsmath}	
\usepackage{amssymb}	
\usepackage{hyperref}
\usepackage{deluxetable}
\usepackage{braket}
\usepackage{amsmath}
\usepackage{bm}
\usepackage{enumitem}
\usepackage{rotating}

\usepackage{newtxtext,newtxmath}

\title[Reionization-Era Quasars observed with FIRE]{Evidence for Episodic Black Hole Growth of Reionization-Era Quasars observed with Magellan/FIRE}

\author[L. Bigwood et al.]{
Leah Bigwood,$^{1,2}$\thanks{E-mail: lmb224@cam.ac.uk}
Anna-Christina Eilers,$^{3}$
Robert A. Simcoe,$^{3}$
\\
$^{1}$ Department of Physics, Durham University, South Road, Durham DH1 3LE, UK
\\
$^{2}$ Institute of Astronomy, Madingley Road, Cambridge, CB3 OHA
\\
$^{3}$
MIT Kavli Institute for Astrophysics and Space Research, 77 Massachusetts Ave., Cambridge, MA 02139, USA\\
}

\date{Accepted XXX. Received YYY; in original form ZZZ}

\pubyear{2024}

\begin{document}
\label{firstpage}
\pagerange{\pageref{firstpage}--\pageref{lastpage}}
\maketitle

\begin{abstract}
Observations of high-redshift quasars hosting billion solar mass black holes at $z\gtrsim6$ challenge our understanding of early supermassive black hole (SMBH) growth. In this work, we conduct a near-infrared spectroscopic study of $19$ quasars at $6.2\lesssim z\lesssim 7.5$, using the Folded-port InfraRed Echellette (FIRE) instrument on the $6.5$-meter Magellan/Baade Telescope. We estimate the single-epoch masses of the quasars' SMBHs by means of
the \mgii\ emission line and find black hole masses of $M_{\text{BH}} \approx(0.2-4.8)\,\times\,10^9\,M_\odot$. 
Furthermore, we measure the sizes of the quasars' proximity zones, which are regions of enhanced transmitted flux bluewards of the \lya emission line, ionized by the quasars' radiation itself.  While it has been shown that the proximity zone sizes correlate with the quasars' lifetimes due to the finite response time of the intergalactic medium to the quasars' radiation, we do not find any correlation between the proximity zone sizes and the black hole mass, which suggests that quasar activity and the concomitant black hole growth are intermittent and episodic. 
\end{abstract}

\newcommand{\angstrom}{\mbox{\normalfont\AA}}


\section{Introduction}

Quasars are the most luminous, non-transient sources in the observable Universe, and thus they have been observed well into the epoch of reionization.  To date more than $200$ quasars 
above a redshift of $z\gtrsim 6$ have been discovered \citep{Fan2022}, with the current redshift-record holder at $z=7.64$ \citep{Wang2021}, which is at a time when the universe is only $\approx 700$~Myr old.  It is well established that their emission is powered by the accretion of material onto a central super-massive black hole (SMBH) from the surrounding accretion disk \citep[e.g.][]{Salpeter1964,Lyden-Bell1969,Kormendy1995,Richstone1998}.  Analyses of the quasars' rest-frame UV and optical spectra, including those discovered in the reionization era, have demonstrated that the quasars host SMBHs with masses exceeding $M_{\text{BH}}\sim 10^9 M_{\odot}$ \citep[e.g.][]{2011Natur.474..616M, 2017ApJ...849...91M, Eilers_2020}. 

The existence of billion solar mass SMBHs at high redshifts challenge our understanding of black hole growth due to the short amount of cosmic time available to reach such masses.  It has been argued that in order to grow such SMBHs from a constant supply of fuelling material, the stellar remnant seeds must have masses of order $M_{\mathrm{BH}}\sim 100$~$M_{\odot}$ and of order $t_{\mathrm{Q}}\sim 10^9$~years of quasar activity, i.e. when the SMBH's active accretion is powering the quasar's UV luminosity, is required, even if accretion is continuous at the Eddington limit \citep{Volonteri2012}.  Given that this time is comparable to the age of the Universe for quasars observed at $z>6$, constraining the timescale that quasars are active is imperative to understanding whether these objects exacerbate the challenges in the theory of SMBH formation and growth.  


However, measuring the timescales of quasar activity has proven to be challenging, even for objects at lower redshift. At $z \sim 2 \mathrm{-}  4$, weak constraints of $t_{\mathrm{Q}} \sim 10^6 \mathrm{-} 10^9$~years have been obtained by means of the spatial clustering of quasars and the inferred abundance of their host dark matter halos. This relies on the idea that rarer objects tend to be highly biased tracers of the underlying dark matter distribution. Therefore, a comparison of the number density of quasars to the abundance of their host dark matter halos provides an estimate of the duty cycle of luminous quasars \citep[e.g.][]{Haiman2001,Martini2001}.  A further study of local early-type galaxies by \citet{Yu2002} constrained $t_{\mathrm{Q}} \sim (3-13) \times 10^7$ years, which was estimated under the assumption that if black hole growth predominantly takes place during optically bright quasar phases, the quasar luminosity function should be reflective of the gas accretion history of local black holes. Spectroscopic studies have also provided additional constraints by detecting changes in quasar luminosities, and therefore ionisation rates, and the time-lag to the resulting change in the intergalactic medium's (IGM) opacity. This method has resulted in estimates in the range $t_{\mathrm{Q}} \sim 10^5 \mathrm{-} 10^7$~years \citep[e.g][]{Adelberger2004}.  Tighter constraints on the timescale of quasar activity, and estimates for the objects at higher redshift which challenge current SMBH formation theory, are clearly necessary to progress our understanding. 

For quasars at $z\gtrsim 5.5$ it has been shown that timescales of their nuclear activity can be estimated by means of the size of their ``proximity zones'' \citep[see, e.g.][]{Eilers_2017,Davies_2019,Eilers2021}.  Quasars ionize the surrounding IGM, which leads to a deficit of neutral hydrogen and therefore reduced Ly$\alpha$ absorption.  This results in a region of enhanced transmitted flux in quasar spectra occurring immediately bluewards of the Ly$\alpha$ emission line.  Due to the finite response time of the IGM to the ionizing radiation, the size of this region has a dependency on the amount of time the quasar has been active. 
At $z\sim 6$ the recombination time of the intergalactic medium to become sufficiently neutral to be opaque to Ly$\alpha$ photons is short, and thus the proximity zone sizes only reflect the time of the last accretion episode.  The situation changes with increasing redshift; at $z\gtrsim7$ the epoch of reionization is not yet complete and therefore the IGM is increasingly neutral.  This complicates the mapping from proximity zone size to lifetime, with the integrated lifetime being measured as opposed to an individual episode.  The spectra of quasars in highly neutral IGM exhibit a damping wing on the \lya\ emission line predominantly due to the intrinsic line width and the velocity dispersion of atoms \citep{Miralda1998,Davies2019,Simcoe2012}.  
The damping wing has thus far only been observed in objects at $z>7$ \citep[e.g.][]{Greig2017,Davies2018c,Yang2020}.  In this work we measure the proximity zone sizes of objects at $z<7$, where we expect the IGM to already be mostly ionized.  As a result, we hereinafter define the `lifetime' of a quasar as the time elapsed since the \textit{current} luminous phase began, since this is the timescale the proximity zone size of $z\lesssim7$ quasars is sensitive to.

Previous studies of the proximity zones of observed reionization era quasars have resulted in lifetime estimates to be as short as $t_{\mathrm{Q}}<10^4$~years \citep{Eilers2021} with an average quasar lifetime of $t_{\mathrm{Q}}\sim 10^6$~years \citep{Morey2021}.  If SMBH growth occurs in a single episode and only when the quasar is optically bright, this poses the question of whether they form from more massive initial seeds, or whether the accretion could happen in a radiatively inefficient manner. The simple exponential light curves in which a quasar emits at a constant luminosity throughout its entire lifetime, so-called ``light-bulb'' light curves, of quasars has also been questioned, and in reality quasars may undergo multiple episodes of luminous quasar activity with concurrent black hole growth, alternated with quiescent phases \citep[e.g.][]{2011ApJ...737...26N,2015MNRAS.451.2517S,Davies2020}.  This behaviour has been demonstrated in hydrodynamical simulations, with episodic quasar lifetimes and intermittent black hole growth being observed in works such as \citet[][]{Novak2011}.  A more complete sample of masses and lifetimes for high redshift quasars is clearly required in order to address these possibilities of their formation and growth.

In this paper we carry out a spectroscopic study of $19$ quasars at $6.17<z<7.51$.  At these redshifts the rest-frame UV emission is shifted to the near-infrared (NIR) wavelengths, hence we utilise the capabilities of the Folded-port InfraRed Echellette (FIRE) instrument on the Magellan/Baade Telescope to obtain NIR spectra of these quasars.  We measure the sizes of proximity zones for the quasars at $z<7$, noting that we leave the modelling of the damping wing signature for the highest redshift objects for a companion paper \citep{Durovcikova2024}.  We also estimate the single epoch masses of the host SMBH of all quasars in the sample by means of the \mgii\ line.  By testing for correlation between these two measurements we suggest that quasar activity is episodic and intermittent, and discuss obscured growth as a resolution to the tension with SMBH formation and growth models. 

Our paper is structured as follows: \S~\ref{sec:quasarsample} introduces the quasar sample and the data reduction procedure. In \S~\ref{sec:masses} we obtain single-epoch SMBH mass estimates, as well as redshifts derived from the \mgii\ line.  We then determine the continuum emission of each quasar in the \lya\ forest in \S~\ref{sec:continuum}, which is required for the proximity zone measurements shown in \S~\ref{sec:proximsize}.  We go on to discuss the implications of our results on quasar lifetimes and SMBH growth in \S~\ref{sec:discuss}. We summarize our findings in \S~\ref{sec:summary}. Throughout the manuscript we assume a flat $\Lambda$CDM cosmology ($h=0.677$, $\Omega_m=0.307$) consistent within the 1$\sigma$ results of \cite{2016A&A...594A..13P}.

\section{Quasar Sample}
\label{sec:quasarsample}
The quasar sample contains $19$ quasars at $6.2\lesssim z\lesssim 7.5$ totalling more than 83 hours of observations with the FIRE instrument on the $6.5$-meter Magellan/Baade Telescope, which were conducted between January 2021 and August 2022. All observations are obtained in Echelle mode using an $0.6\arcsec$ slit. The details of the observations are shown in Table~\ref{tab:obs}. 


\begin{table*}
\caption{Observational information for the quasars in our sample, as well as the best redshift estimate available in literature.  We also list the absolute magnitudes $M_{1450}$ for the subset of quasars for which this is required to correct the proximity zone measurement (see \S~\ref{sec:proximsize}).} 

\begin{tabular}{ccccccc}
\hline
Object & $z$ & $M_{1450}$  & Date & $t_{\mathrm{exp}}$~[s] \\
\hline
J0020--3653 & 6.834 $\pm$ 0.001$^{\rm{d}}$ & -26.92 & Jan 2021 & 3600  \\
&& & Jun 2021 & 14400  \\ 
& && Oct 2021 & 10800\\
J0216--5226 & 6.41$\pm$0.05$^{\rm{h}}$ & - &  Jan 2021 & 9600 \\
J0218+0007 & 6.7700 $\pm$ 0.0013$^{\rm{c}}$& -25.55 & Jan 2021 & 7200  \\
&& & Dec 2021 & 9000 \\
J0244--5008 & 6.724$\pm$0.001$^{\rm{b}}$   & -26.72& Jan 2021 & 4800\\
&& & Sep 2021 & 7200 \\
&& & Oct 2021 & 10800 \\
&& & Jan 2022 & 9000 \\
J0252--0503 & 7.0006$\pm$ 0.0009$^{\rm{c}}$& - & Jan 2021 & 5400  \\
&& & Sep 2021 & 2400 \\
J036+03 & 6.5412$\pm$0.0018$^{\rm{f}}$&-27.4 & Aug 2022 & 7200 \\
J0319--1008 & 6.8275 $\pm$ 0.0021$^{\rm{c}}$ & -25.36& Oct 2021 & 16200 \\
J0410--4414  & 6.21$\pm$0.01$^{\rm{b}}$ & -26.14& Jan 2021 & 4800\\
J0411--0907 & 6.8260 $\pm$ 0.0007$^{\rm{c}}$& -26.58 & Nov 2021 & 15600 \\
J0525--2406 & 6.5397 $\pm$ 0.0001$^{\rm{c}}$ & -25.47 & Dec 2021 & 10800\\
J0910+1656 & 6.7289 $\pm$ 0.0005$^{\rm{c}}$& -25.34 & Mar 2021 & 12600   \\
J0921+0007 & 6.5646 $\pm$ 0.0003$^{\rm{c}}$ & -25.19& Jan 2021 & 7200  \\
& && Apr 2021 & 4800 \\
J1007+2115 & 7.5149 $\pm$ 0.0004$^{\rm{c}}$& - & Jan 2021 & 14400  \\
&& & Mar 2021 & 7200 \\
&& & Apr 2021 & 7200 \\
J1104+2134 & 6.7662 $\pm$ 0.0009$^{\rm{c}}$ & -26.63& Jan 2021 & 9600  \\ 
 && & Mar 2021 & 5400 \\
& & & Apr 2021 & 9600 \\
& & & Jun 2021 & 6000\\
J1217+0131 & 6.17$\pm$ 0.05$^{\rm{e}}$ & -25.76& Jun 2021 & 13200 \\
J1526--2050 & 6.5864 $\pm$0.0005$^{\rm{a}}$ & -27.20& Jul 2021 & 9600  \\
&& & Apr 2022 & 13800 \\
J1535+1943 & 6.370 $\pm$ 0.001$^{\rm{c}}$& - & Apr 2021 & 8400  \\
J2002--3013 & 6.6876 $\pm$ 0.0004$^{\rm{c}}$ & -26.9& Mar 2021 & 10200\\
&& & Apr 2021 & 4800 \\
& & & Jul 2021 & 4200\\
J2102--1458 & 6.6645 $\pm$ 0.0002$^{\rm{c}}$ & -25.53 & Apr 2021 & 4800 \\
\hline
\end{tabular}
\label{tab:obs}
\tablecomments{a: \citet{Decarli2018}; b: \citet{Reed2017}; c: \citet{Yang2021}; d: \citet{reed2019}; e: \citet{Wang2017}; f: \citet{Banados2015}; g: Eduardo et al. (in prep.); h: \citet{Yang2019}}

\end{table*}

The spectra are homogeneously reduced using the open-source python-based spectroscopic data reduction pipeline \texttt{PypeIt}\footnote{\url{https://github.com/pypeit/PypeIt}} version 1.7.1 \citep{Pypeit}. The details of this pipeline are described in \citet{Prochaska2020}, 
but we will summarize them here briefly. In short, we obtain an on-sky wavelength solution based on sky OH lines, and perform the sky subtraction on the 2D images by differencing exposures dithered along the slit. We fit a $b$-spline to further eliminate sky line residuals following \citet{Bochanski_2009}. We then perform an optimal extraction \citep{Horne1986} to obtain the 1D spectra. All spectra are flux calibrated using sensitivity functions of standard stars, ideally observed during the same night. 

We co-add the flux-calibrated 1D spectra from each night and correct for telluric absorption features by jointly fitting an atmospheric model and a quasar model. The telluric model grids are produced using the Line-By-Line Radiative Transfer Model \citep[LBLRTM;][]{Clough2005}. 

All reduced spectra are shown in Figure~\ref{fig:continua}. 

The best published redshift estimates for the majority of these quasars are derived from the \cii\  157.74 $\mu$m emission line.  When this is not available, the systematic redshifts are traced by the \mgii\ emission line or the extended \lya\ emission.  
Table~\ref{tab:obs} lists the best redshift estimates in the literature. 


\section{Analysis}
\subsection{Black Hole Mass Estimates}
\label{sec:masses}
Under the assumption that the gravitational pull of the central black hole dominates the dynamics in the broad line region (BLR) of the quasar, we can apply the virial theorem. 
This allows an estimation of the black hole's mass to be obtained by means of the width of the broad emission lines arising from the gas in the BLR and the luminosity output.  We utilise the black hole mass scaling relation of \citet{2009ApJ...699..800V}, i.e.
\begin{equation}
    \frac{M_{\text{BH}}}{M_{\odot}}=10^{6.86}\left(\frac{\text{FWHM}_{\text{MgII}}}{10^3 \text{kms }^{-1}}\right)^2\left(\frac{\lambda L_{\lambda,3000 \angstrom}}{10^{44}\text{ erg s}^{-1}}\right)^{0.5}, 
\end{equation}
which was calibrated to the existing \hb \ and \civ \ scaling relations using several thousand high-quality quasar spectra from  Sloan Digital Sky Survey Data Release 3 (SDSS DR3) quasar sample ranging $0.08<z<5.41$.  Here FWHM$_{\text{MgII}}$ is the full width at half maximum (FWHM) of the \mgii\ emission line and $L_{\lambda, 3000 \angstrom}$ is the monochromatic luminosity at $\lambda=3000$~\AA.  The intrinsic $1\sigma$ scatter in the scaling relation is approximately 0.55 dex. 

We model the quasar emission in the spectral region $2100\leq \lambda_{\text{rest}}\leq3088$~\AA\  as a superposition of a power law continuum with slope $\alpha$ arising from the accretion disk, a scaled template spectrum of the iron emission lines \ion{Fe}{ii} and \ion{Fe}{iii} in the BLR, as well as a single Gaussian of variance $\sigma^2_{\mgii}$ centred on $\mu_{\mgii}$ to model the \mgii\ line, i.e.
\begin{equation}
    f_{\lambda} = a_0\cdot \left(\frac{\lambda}{2500}\right)^{-\alpha}+a_1\cdot f_{\lambda,\text{iron}} + a_2\cdot \exp\left[{-\frac{(\lambda - \mu_{\mgii})^2}{2\sigma^2_{\mgii}}}\right]
\end{equation}
where $a_{0}$, $a_{1}$ and $a_{2}$ are the amplitudes of the individual components.  We use the iron template spectrum of \citet{Tsuzuki2006}, which was derived from a narrow emission line quasar.  We note that we utilise the template of \citet{Tsuzuki2006} over \citet{Vestergaard2001},  since the former separates the \mgii\ emission from the underlying \feii\ emission, whereas the latter sets the \feii\ emission to zero in the region of the \mgii\ line.  This resulted in the model of the quasar emission around the \mgii\ line showing a better quality of fit to the spectra in our sample.  We however caveat that the \citet{2009ApJ...699..800V} scaling relation was constructed using the iron template of \citet{Vestergaard2001}, and that \citet{Schindler2020} demonstrated that using the \citet{Tsuzuki2006} template with this scaling relation can bias SMBH masses low by $\sim0.21$~dex.  To mimic the broad emission lines present in our quasars' spectra, we convolve the iron template with a Gaussian kernel of width FWHM $\approx$ {FWHM}$_{\mgii}$.  Prior to fitting the model, we rescale the reduced spectrum so that the integrated flux within the J-band is in agreement with the  reported magnitude J-band magnitude in literature, if available. 

\begin{figure*}

    \centering
	\includegraphics[width=1.85\columnwidth]{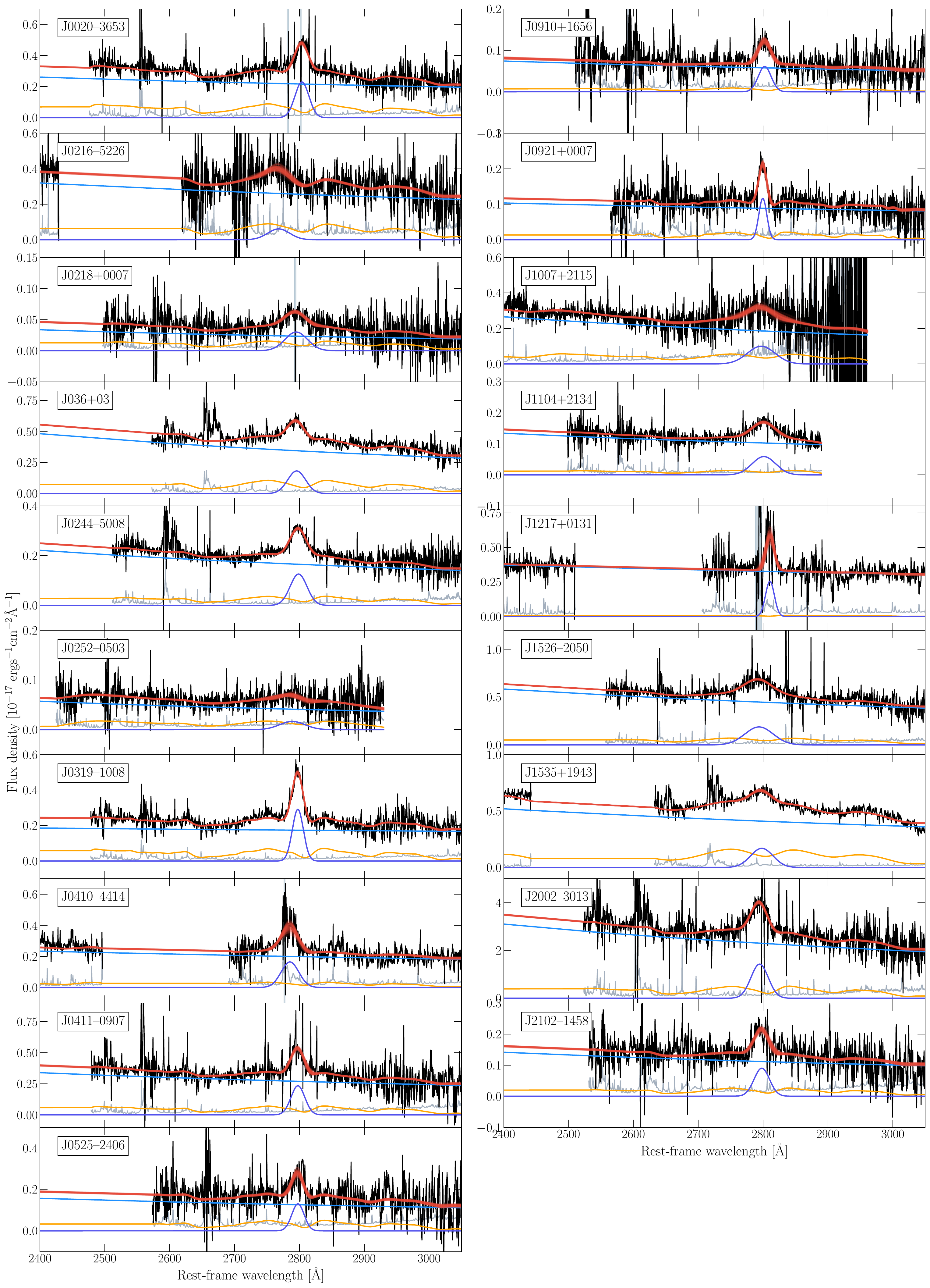}
    \caption{The multi-component fits to the spectral region around the MgII emission line, shown in the quasars rest-frame using the best redshift estimate.  The spectrum and its noise vector have been smoothed with a boxcar function of width ten pixels, and are shown in black and grey respectively.  Regions of strong telluric absorption were removed, and further masking was applied to broad absorption lines and noisy spikes, shown by the shaded grey areas.  The faint red curves show 100 draws from the posterior distribution of the best-estimate model.  The individual components are shown as the colored curves: a power-law continuum (blue curve), the iron template spectrum broadened to mimic the emission from the BLR (yellow curve), and a Gaussian to model the \mgii\ emission line (purple curve).
    }
    \label{fig:masses}
\end{figure*}

By means of the Markov chain Monte Carlo sampler \texttt{emcee} \citep{2013PASP..125..306F}, we estimate the free parameters of the model assuming flat priors
and adopt the mean of the posterior probability distribution as the best parameter estimate. The monochromatic luminosity $L_{\lambda, 3000}$ was determined by estimating the flux at a rest-fram wavelength of $\lambda=3000$~\AA\ from the best-fit power law continuum.  The best fits to the \mgii\ emission lines for each quasar in our sample are displayed in Figure \ref{fig:masses}, and the corresponding mass estimates are listed in Table~\ref{tab:results}.  

We measure the quasar redshift derived from the \mgii\ emission line $z_{\mgii}$ using the estimations of the Gaussian centre, $\mu_{\mgii}$.  Note that this emission does not arise from the host galaxy (unlike the \cii\ line), and instead originates in the BLR.  As a result, the emission may experience modest velocity shifts with respect to the systematic redshift of the quasar.  The average shift with respect to the quasar's systematic redshift is $\Delta v(\mgii\ - \cii\ ) = 391$ kms$^{-1}$ \citep{Schindler2020}.  
The $z_{\mgii}$ estimates are listed in Table~\ref{tab:results}.  We note that we measure the \mgii\ derived redshift of J0216-5226 to be $z_{\mgii}=6.3283\pm0.0094$, which is blueshifted from the best available literature estimate of $z=6.41\pm0.05$ \citep{Yang2019} by $\sim 3305$ kms$^{-1}$.  \citet{Yang2019} attained the measurement by matching quasar template spectra to the \lya and \nv \ emission lines, noting that the offset of the \lya line with respect to the systematic redshift is typically $\Delta v( \rm{Ly}\alpha\ - \cii\ ) \sim 500$ kms$^{-1}$.  We therefore find that the estimated $z_{\mgii}$ for this object exhibits a blueshift in significant excess to the typical systematic velocity offset.



\begin{figure}

    \centering
	\includegraphics[width=\columnwidth]{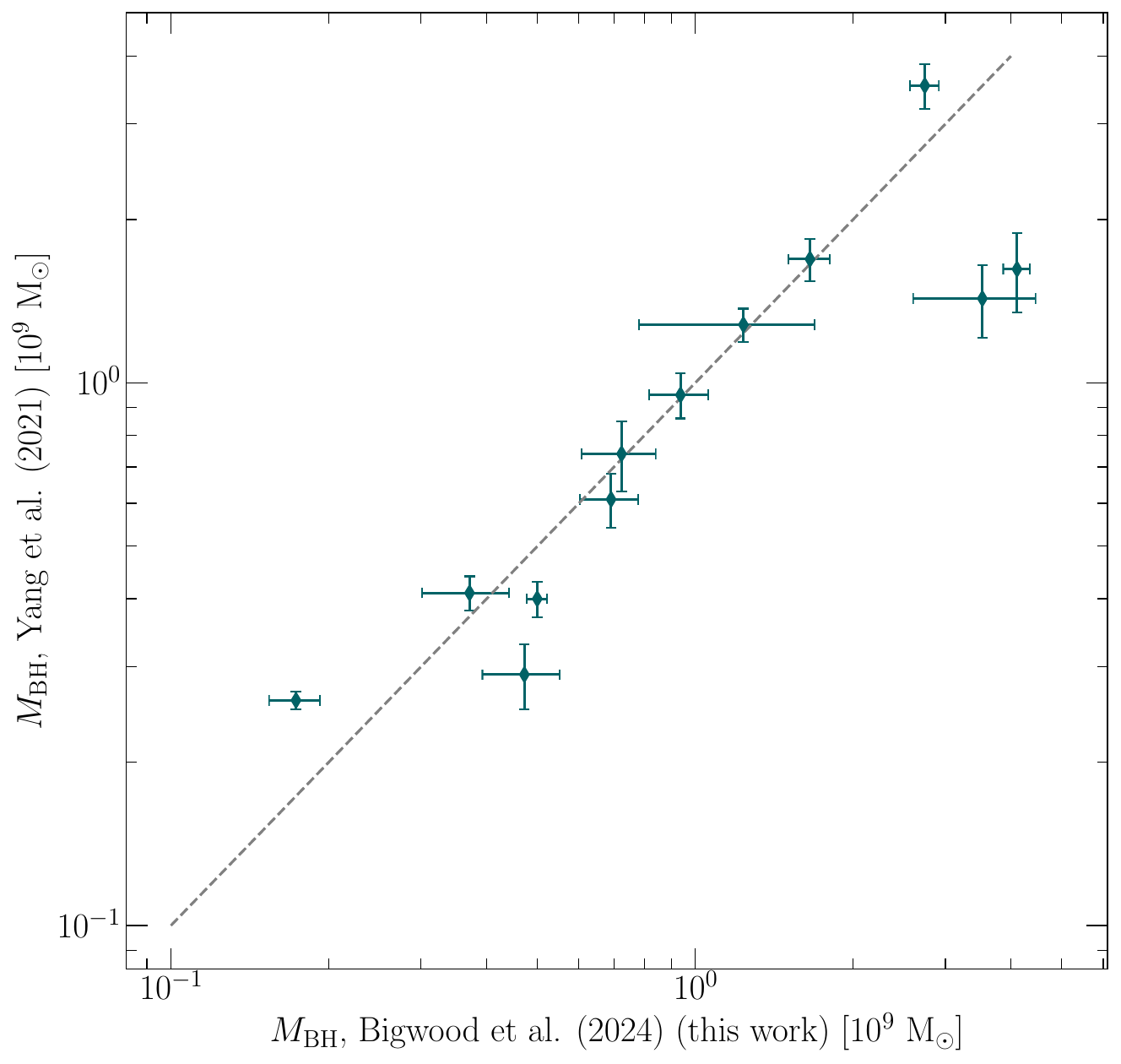}
    \caption{Comparing the masses of host SMBHs estimated in this paper to those obtained in \citet{Yang2021}, for the quasars that are characterised in both works.  The errorbars displayed are purely statistical.  There is an intrinsic systematic scatter in $M_{\mathrm{BH}}$ of approximately 0.55 dex arising from the scaling relation of \citet{2009ApJ...699..800V}.
    }
    \label{fig:mass_compare}
\end{figure}

Several quasars in our sample overlap with those characterised previously in \citet{Yang2021}. 
 Figure~\ref{fig:mass_compare} compares the respective mass estimates obtained, finding overall a reasonable agreement between the two studies within the intrinsic scatter of the scaling relation. 

\subsection{Quasar Continuum}
\label{sec:continuum}
In order to measure the proximity zone we first require an estimate for the quasar continuum emission in the \lya\ forest. Residual neutral hydrogen present in the IGM leads to significant absorption bluewards of the \lya\ emission line for the spectra of high redshift quasars.  

We obtain the continuum estimate using a Principal Component Analysis \citep[PCA; e.g.][]{Suzuki_2005, Paris2011, Davies_2018}. 
Due to the absorption bluwards of \lya, we only fit to the `red' side of the spectra ($1220<\lambda_{\text{rest}}<2850$~\AA ), and then predict the continuum emission on the `blue' side ($1181<\lambda_{\text{rest}}<1220$~\AA ).  The `red' and `blue' PCA components, $R_i$ and $B_i$, are a set of ten and six basis spectra respectively, obtained with a PCA decomposition of 12764 training spectra from the Sloan Digital Sky Survey Baryon Oscillation Spectroscopic Survey (SDSS BOSS) sample into an orthogonal basis \citep{Davies_2018}. 

We obtain the rest-frame quasar spectra using the best available redshift estimate, i.e. taking those derived from the \mgii\ emission line in \S~\ref{sec:masses} if it improves upon the best previously published redshift estimate. We mask spectral regions with significant atmospheric absorption, and normalise the spectra to unity in a region free of significant emission lines at $1290\pm2.5$~\AA. 
To remove strong metal absorption lines from intervening absorption systems along our line-of-sight we then fit a spline to the `red' region of the spectra, and remove outlying values by sigma clipping at the 3$\sigma$ level. 
We use \texttt{emcee} to estimate the 10 PCA coefficients for these components in the `red' spectral region, $r_i$, thus estimating the quasar continuum in the logarithmic flux space as
\begin{equation}
    \log f_{\lambda}\approx \langle \log f_{\lambda} \rangle + \sum_{i=1}^{10} r_i R_i
\end{equation}
where $\langle \log f_{\lambda} \rangle$ is the mean logarithmic flux. We also fit for a redshift offset $\Delta z$, to account for offsets between the broad emission lines and the systemic redshift. We chose flat priors for each free parameter, taking the $5\sigma$ region for $r_i$ from the training set data \citep[see][for details]{Davies_2018} and $\Delta z\in[-0.15,0.03]$. 
We adopt the mean of the resulting posterior probability distribution as the best parameter estimate.  

\begin{figure*}

    \centering
	\includegraphics[width=1.85\columnwidth]{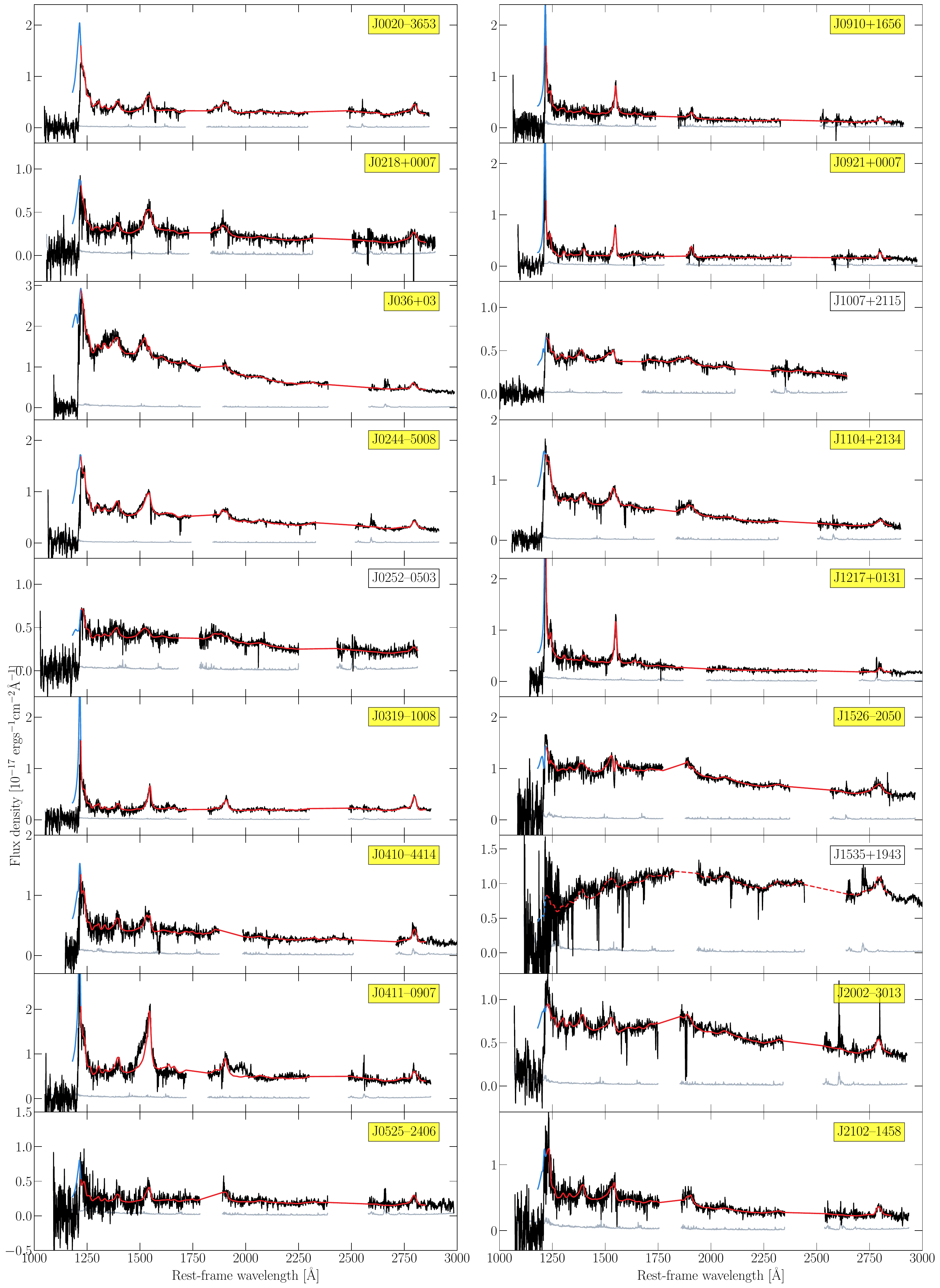}
    \caption{The spectra and best-fit continua model of the quasars in our sample.  We estimate the continua across the full spectral range by fitting 10 `red' principal components to the region $1220<\lambda_{\text{rest}}<2850$~\AA\ (red curve), and projecting onto a set of six `blue' principal components covering $1181<\lambda_{\text{rest}}<1220$~\AA\ (blue curve).  A boxcar smoothing of 20 pixels has been applied to the spectrum (black) and its noise vector (grey).  We have removed regions of strong telluric absorption, and further masking was applied to broad absorption lines and regions of high noise, shown by the shaded grey areas.  Labels in yellow denote objects for which we use the continuum reconstruction in the proximity zone analysis of \S~\ref{sec:proximsize}.  For these objects we enlarge the region showing the continuity between the red and blue continuum reconstruction components in Figure~\ref{fig:continua_zoom}.
    We show the continua model of J1535+1943 as a dashed curve due to its highly usual slope resulting from dust obscuration.  
    }
    \label{fig:continua}
\end{figure*}

The best-estimate $r_i$ are then projected onto the coefficients of the blue PCA components, $b_j$, using a projection matrix $P_{i,j}$ also determined from the training spectra, i.e. 
\begin{equation}
    b_j = \sum_{i=0}^{10}r_i P_{i,j}. 
\end{equation}
We thus obtain a fit for the quasar continua over the entire spectral region $1181<\lambda_{\text{rest}}<2850$~\AA.  Figure \ref{fig:continua} present the continua estimates for the quasars in our sample.  We note that due to the lack of emission lines in the spectra of J0216–5226 we cannot securely determine its continuum, and therefore omit J0216–5226 from the remainder of the analysis.  In Figure~\ref{fig:continua_zoom} we enlarge the region showing the continuity of the red and blue continuum reconstruction components in order to better demonstrate the quality of the continuum reconstruction.  

We note that J1535+1943 is heavily dust obscured, leading to an unusual continuum slope. Hence, we caution that the continuum estimate might be biased due to the lack of similar objects in the PCA training set.  It would be interesting to measure the proximity zone size and infer the quasar lifetime for this obscured object, however simultaneous modelling of the quasar continuum and the obscuration rate would be required.  We therefore do not take this object into account in the proximity zone analysis of the following section and leave a full study of it for future work.

\subsection{Measuring Proximity Zone Sizes}
\label{sec:proximsize}
\begin{figure*}

    \centering
	\includegraphics[width=1.9\columnwidth]{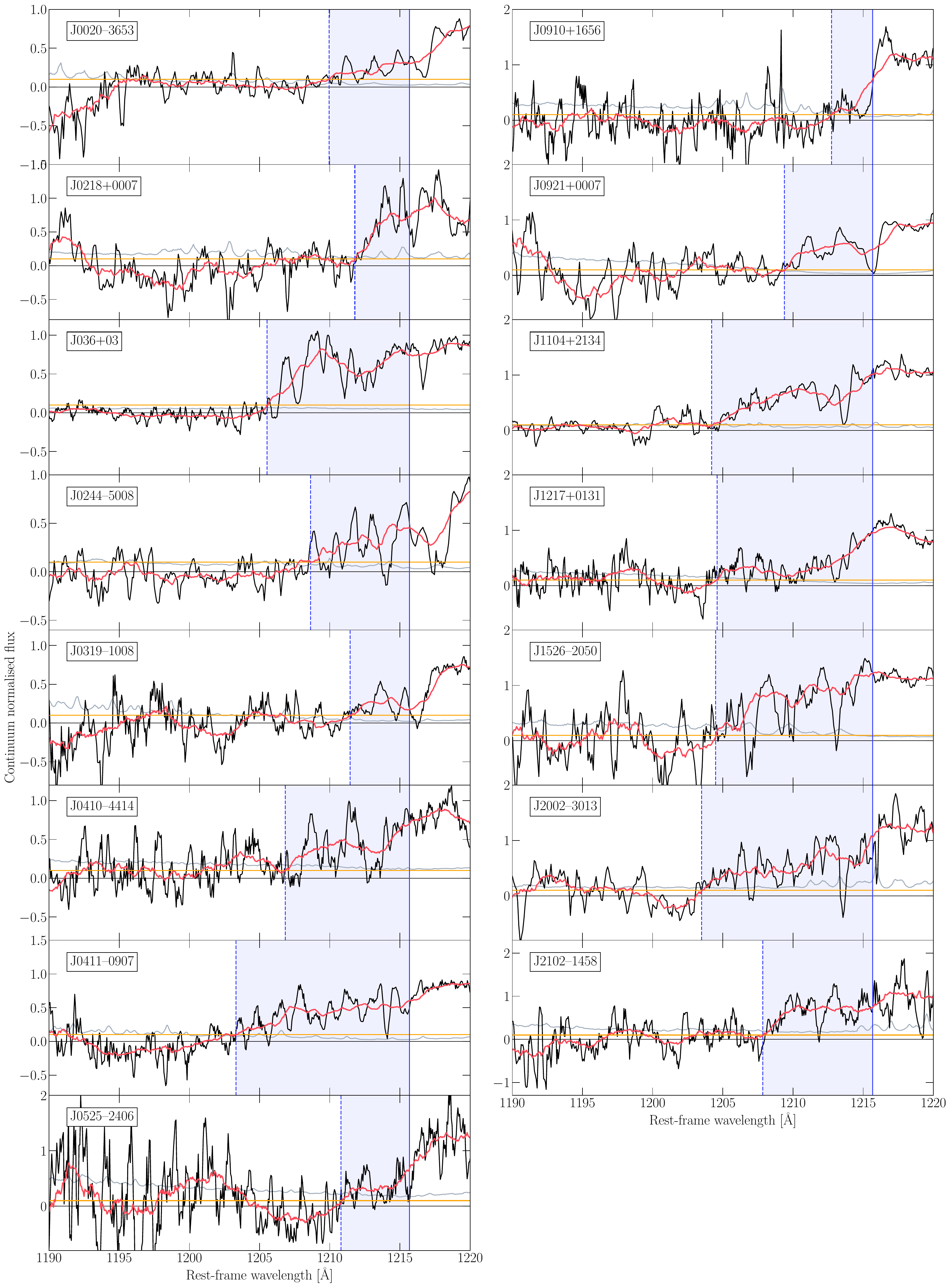}
    \caption{The continuum-normalised spectra and proximity zones of the quasar dataset.  The black and grey curves display the flux and its noise vector respectively, both smoothed with a boxcar function of width 5 pixels.  The red curve shows the continuum-normalised spectra smoothed with a boxcar function of width 20~\AA.  As per the definition adopted in literature, the proximity zone extent is where this curve drops below the 10\% flux level (yellow horizontal line). The proximity zone is shown as the shaded blue region, with the \lya\ emission line shown as the solid blue vertical line and the proximity zone edge the dashed blue vertical line.} 
    \label{fig:proxim}
\end{figure*}

We adopt the standard definition of the proximity zone used in previous studies \citep[e.g.][]{Fan2006, Carilli2010, Eilers_2020, Satyavolu2023}. 
We normalise the rest-frame quasar spectra by the continuum estimate, and apply a boxcar smoothing to the normalised flux of width 20~\AA\ in the observed wavelength frame. At $z=6$ this smoothing scale corresponds to approximately $1$~pMpc or $700~\rm km\,s^{-1}$. The extent of the proximity zone, $R_{\mathrm{p}}$, is defined as the region bluewards of the \lya\ emission line before the smoothed transmitted flux drops below $10\%$.  This follows the standard definition first adopted by \citet{Fan2006}, and utilised in other previous studies \citep[e.g.][]{Willott2007,2017ApJ...849...91M, Eilers_2017}. 
 In Figure~\ref{fig:proxim} we present the smoothed continuum-normalised spectrum and show the extents of the proximity zone sizes for our quasar sample. Note that damping wings in quasar spectra have been detected in objects at $z>7$, imprinted by the high neutral fraction of the IGM \citep[][]{Wang2020, Yang2020}. 
As a result we exclude proximity zone estimates for objects at $z>7$, since they would 
require careful modeling of the neutral gas fraction in the proximity zone size, as shown in previous work \citep[e.g.][]{Davies_2019, Greig2017}. 

The extents of the proximity zones are dependent on the luminosity of the quasar, since a more luminous quasar will create more ionising radiation. Thus, we also calculate the luminosity-corrected proximity zone size, $R_{\rm p,\,corr}$, which has been shown to correlate with the quasar's lifetime \citep[e.g.][]{Eilers_2017, Eilers_2020, Morey2021}. We adopt the relation to eliminate the luminosity dependency introduced in \citet{Eilers_2017}; 
\begin{equation}
    R_{\rm p,\,corr}=R_{\rm p}\times10^{-0.4(-27-M_{1450})/2.35}
\end{equation}
which normalises all proximity zone measurements to an absolute magnitude of $M_{1450}=-27$. Table~\ref{tab:results} records both the measured and magnitude-corrected proximity zone sizes for our set of quasars.

\begin{table*}
\caption{The columns show the name of the quasar, the best published redshift $z$, the measured uncorrected and magnitude-corrected proximity zone sizes, $R_{\rm p}$ and $R_{\rm p,\,{corr}}$, redshift estimates based on the \mgii\ emission line $z_{\mgii}$, the FWHM of the Gaussian fit to the \mgii\ emission line FWHM$_{\text{MgII}}$, the monochromatic luminosity $L_{\lambda, 3000}$, and the SMBH mass estimates $M_{\mathrm{BH}}$.}
\begin{tabular}{cccccccc}
\hline
Object & $z$ & $R_{\mathrm{p}}$ [pMpc] & $R_{\rm p,\,{corr}}$ [pMpc] & $z_{\mgii}$&FWHM$_{\text{MgII}}$ [\AA] &$L_{\lambda, 3000}$ [$10^{43}$ erg s$^{-1}$\AA$^{-1}$] &$M_{\mathrm{BH}}$ [$10^9$ $M_\odot$] \\
\hline
J0020–3653$^{\dag}$ & 6.8340 $\pm$ 0.0010 & 1.70 $\pm$ 0.25 & 1.75 $\pm$ 0.26 & 6.8485 $\pm$ 0.0007 & 2727 $\pm$ 80 & 0.883 $\pm$ 0.008 & 0.88 $\pm$ 0.05 \\
J0216–5226 & 6.4100 $\pm$ 0.0500 & - & - & 6.3288 $\pm$ 0.0098 & 4260 $\pm$ 936 & 0.837 $\pm$ 0.014 & 2.08 $\pm$ 0.92 \\
J0218+0007 &  6.7700 $\pm$ 0.0013 & 1.17 $\pm$ 0.48 & 2.07 $\pm$ 0.84 & 6.7600 $\pm$ 0.0028 & 4350 $\pm$ 273 & 0.085 $\pm$ 0.003 & 0.69 $\pm$ 0.09 \\
J0244–5008$^{\dag}$ &  6.7240 $\pm$ 0.0010 &  2.14 $\pm$ 0.28 & 2.39 $\pm$ 0.31& 6.7240 $\pm$ 0.0008 & 3272 $\pm$ 78 & 0.609 $\pm$ 0.005 & 1.05 $\pm$ 0.05 \\
J0252–0503 &  7.0006 $\pm$ 0.0009 &  -  &  - & 6.9732 $\pm$ 0.0099 & 4905 $\pm$ 902 & 0.167 $\pm$ 0.006 & 1.24 $\pm$ 0.46 \\
J0319–1008$^{\dag}$ & 6.8275 $\pm$ 0.0021 &  1.26 $\pm$ 0.16 & 2.39 $\pm$ 0.31 & 6.8246 $\pm$ 0.0005 & 2150 $\pm$ 47 & 0.742 $\pm$ 0.008 & 0.50 $\pm$ 0.02 \\
J036+03$^{\dag}$ &  6.5412 $\pm$ 0.0018 &3.20 $\pm$  0.47 & 2.74 $\pm$  0.40 & 6.5334 $\pm$ 0.0012 & 3480 $\pm$ 118 & 1.135 $\pm$ 0.009 & 1.62 $\pm$ 0.11 \\
J0410–4414$^{\dag}$  & 6.2100 $\pm$ 0.0100 & 3.00 $\pm$  0.92 & 4.21 $\pm$  1.29  & 6.1759 $\pm$ 0.0022 & 3439 $\pm$ 332 & 0.602 $\pm$ 0.008 & 1.15 $\pm$ 0.22 \\
J0411–0907 & 6.8260 $\pm$ 0.0007 &  3.68 $\pm$  0.25 & 4.34 $\pm$ 0.30 & 6.8233 $\pm$ 0.0016 & 2686 $\pm$ 174 & 1.071 $\pm$ 0.015 & 0.94 $\pm$ 0.12 \\
J0525–2406  & 6.5397 $\pm$ 0.0001 & 1.53 $\pm$ 0.04 & 2.79 $\pm$ 0.07 & 6.5372 $\pm$ 0.0022 & 2384 $\pm$ 200 & 0.438 $\pm$ 0.012 & 0.47 $\pm$ 0.08 \\
J0910+1656 & 6.7289 $\pm$ 0.0005 & 0.89 $\pm$ 0.18 & 1.71 $\pm$ 0.35 & 6.7391 $\pm$ 0.0024 & 2517 $\pm$ 236 & 0.218 $\pm$ 0.008 & 0.37 $\pm$ 0.07 \\
J0921+0007 &  6.5646 $\pm$ 0.0003 & 1.97 $\pm$ 0.11 & 4.01 $\pm$ 0.23 & 6.5669 $\pm$ 0.0008 & 1561 $\pm$ 87 & 0.321 $\pm$ 0.006 & 0.17 $\pm$ 0.02 \\
J1007+2115   & 7.5149 $\pm$ 0.0004 &  -  &  -  & 7.5099 $\pm$ 0.0079 & 5376 $\pm$ 704 & 0.947 $\pm$ 0.025 & 3.53 $\pm$ 0.93 \\
J1104+2134  & 6.7662 $\pm$ 0.0009 & 3.46 $\pm$ 0.33 & 4.00 $\pm$ 0.38 & 6.7731 $\pm$ 0.0018 & 4592 $\pm$ 205 & 0.391 $\pm$ 0.005 & 1.65 $\pm$ 0.15 \\
J1217+0131$^{\dag}$  & 6.1700 $\pm$ 0.0500 & 3.74 $\pm$ 0.88 & 6.08 $\pm$ 1.43 & 6.2002 $\pm$ 0.0021 & 1717 $\pm$ 186 & 0.981 $\pm$ 0.009 & 0.37 $\pm$ 0.08 \\
J1526–2050  & 6.5864 $\pm$ 0.0005 & 3.50 $\pm$ 0.19 & 3.23 $\pm$ 0.18 & 6.5729 $\pm$ 0.0014 & 5530 $\pm$136 & 1.559 $\pm$ 0.010 & 4.79 $\pm$ 0.24 \\
J1535+1943 &   6.3700 $\pm$ 0.0010 &  - &  -  & 6.3681 $\pm$ 0.0014 & 4375 $\pm$ 140 & 1.301 $\pm$ 0.006 & 2.74 $\pm$ 0.18 \\
J2002–3013  & 6.6876 $\pm$ 0.0004 &  3.73 $\pm$ 0.15 & 3.88 $\pm$ 0.15 & 6.6764 $\pm$ 0.0011 & 3373 $\pm$ 99 & 8.285 $\pm$ 0.073 & 4.11 $\pm$0.24 \\
J2102–1458  & 6.6645 $\pm$ 0.0002 & 2.41 $\pm$  0.07 & 4.28 $\pm$ 0.13 & 6.6625 $\pm$ 0.0027 & 3001 $\pm$ 241 & 0.410 $\pm$ 0.009 & 0.72 $\pm$ 0.12 \\
\hline
\end{tabular}
\label{tab:results}
\tablecomments{\dag: Objects for which the redshift estimate based on the \mgii\ emission line, $z_{\mgii}$, derived in this work is an improvement on the best published redshift $z$.}
\end{table*}

\section{Discussion}
\label{sec:discuss}

To date, over 100 \mgii-based SMBH mass estimates have been obtained for quasars at $z>5.9$ \citep{Fan2022}.  Previous studies such as \citet{Yang2021}, \citet{Farina2022} and \citet{Mazzucchelli2023} report SMBH masses in the range $\sim(0.3-12.6) \times 10^9 \,M_{\odot}$ for quasars observed in the late stages of cosmic reionization.  The mass estimates derived in this work lie within $M_{\text{BH}} \approx(0.2-4.8)\,\times\,10^9\,M_\odot$, and are thus broadly consistent with those reported in the literature. We note that our sample does not include some of the heaviest SMBHs reported in the literature exceeding $M_{\text{BH}} \gtrsim 10^{10}\,M_\odot$ \citep{Wu2015, Mazzucchelli2023}.

For quasars at $z>5.7$, proximity zone sizes have thus far been measured in 87 objects.  \citet{Satyavolu2023} recently presented a study of 22 quasars at $5.8<z<6.6$ and inferred proximity zones in the range $R_{\rm p}\sim 2-7$~pMpc.  Work by \citet{Eilers_2017} discovered three quasars with exceptionally small proximity zones of $R_{\rm p,\,corr}< 2$~pMpc, with \citet{Eilers_2020} finding a further four objects which meet this criteria.  After considering patchy reionization and proximate absorbers as causes for a potential truncation of the proximity zone, they conclude that the most compelling explanation for the extremely small proximity zone sizes are quasar lifetimes as short as $t_{\mathrm{Q}} < 10^5$ years.  In this study, we measure luminosity corrected proximity zones in the range $R_{\rm p,\,corr}\approx 1.7-6.1$~pMpc, consistent with the previous measurements in literature.  We find two objects (J0020–3653 and J0910+1656) with small proximity zones of $R_{\rm p,\,corr}< 2$~pMpc, potentially indicative of very young objects. In a companion paper we obtain precise lifetime estimates for the quasars in our sample \citep{Durovcikova2024} to study their SMBH growth phases.

It is known that the size of proximity zones are correlated with quasars' UV luminous lifetimes, during which the black hole growth occurs \citep[e.g.][]{Salpeter1964, Soltan1982, Eilers_2017, Khrykin2016, Davies2020}.  With this in mind, if quasars were to have only one accretion episode, i.e. a duty cycle of unity, and similar initial black hole seeds, we would expect to find a correlation between proximity zone size and black hole mass. In Figure~\ref{fig:massvsproxim} we test this scenario by showing the estimated SMBH masses against the luminosity corrected proximity zone sizes.  We find a Pearson correlation coefficient of $r=0.054$, indicating no correlation between the two measurements. 
This suggests that SMBHs likely grow in multiple accretion episodes and that quasar light curves do not follow a simple light-bulb model, as has been suggested in previous studies \citep[e.g.][]{Novak2011,Schawinski2015,Davies2020,Satyavolu2023a}. 
However, short accretion episodes as implied by the small proximity zones measured in quasars at $z\approx6$ \citep[e.g.][]{Eilers2021} do not alleviate, and in fact aggravate, the challenge of growing billion solar mass black holes by $z\sim 6$.
Recently, \citet{Satyavolu2023a} proposed the solution that quasars may undergo phases of obscured growth, when the black hole grows without emitting optical radiation along a given sightline.  By allowing black hole growth when the quasar isn't optically bright, they demonstrated that super-Eddington accretion is not necessarily a requirement of achieving the black hole masses we observe at high redshift.  This behaviour has also been displayed in hydrodynamical simulations, with \citet{Bennett2023} showing that simulated quasars hosting $\sim10^9 M_{\odot}$ black holes at $z=6$ spend a significant amount of time with sightlines obscured in the UV, as a result of surrounding dust and gas.  Obscured growth therefore provides a potential reconciliation between the short proximity zone sizes and host SMBH masses measured for high redshift quasars, thus aiding the tension in black hole formation and growth models. 

It is important to note that our conclusions rely on the correlation between proximity zone size and lifetime, which assumes that the quasars we measure at $z<7$ reside in an ionised IGM.  Although we do not detect the damping wing signature of a neutral IGM in our sample, we cannot rule out the possibility that patches of neutral hydrogen are still present around the quasars at $z>6$.  Future work will however account for this by modelling the reionization of the IGM and attaining robust measurements of quasar lifetimes, better informing us on the theories we have suggested.






\begin{figure}
    \label{fig:massvsproxim}
    \centering
	\includegraphics[width=\columnwidth]{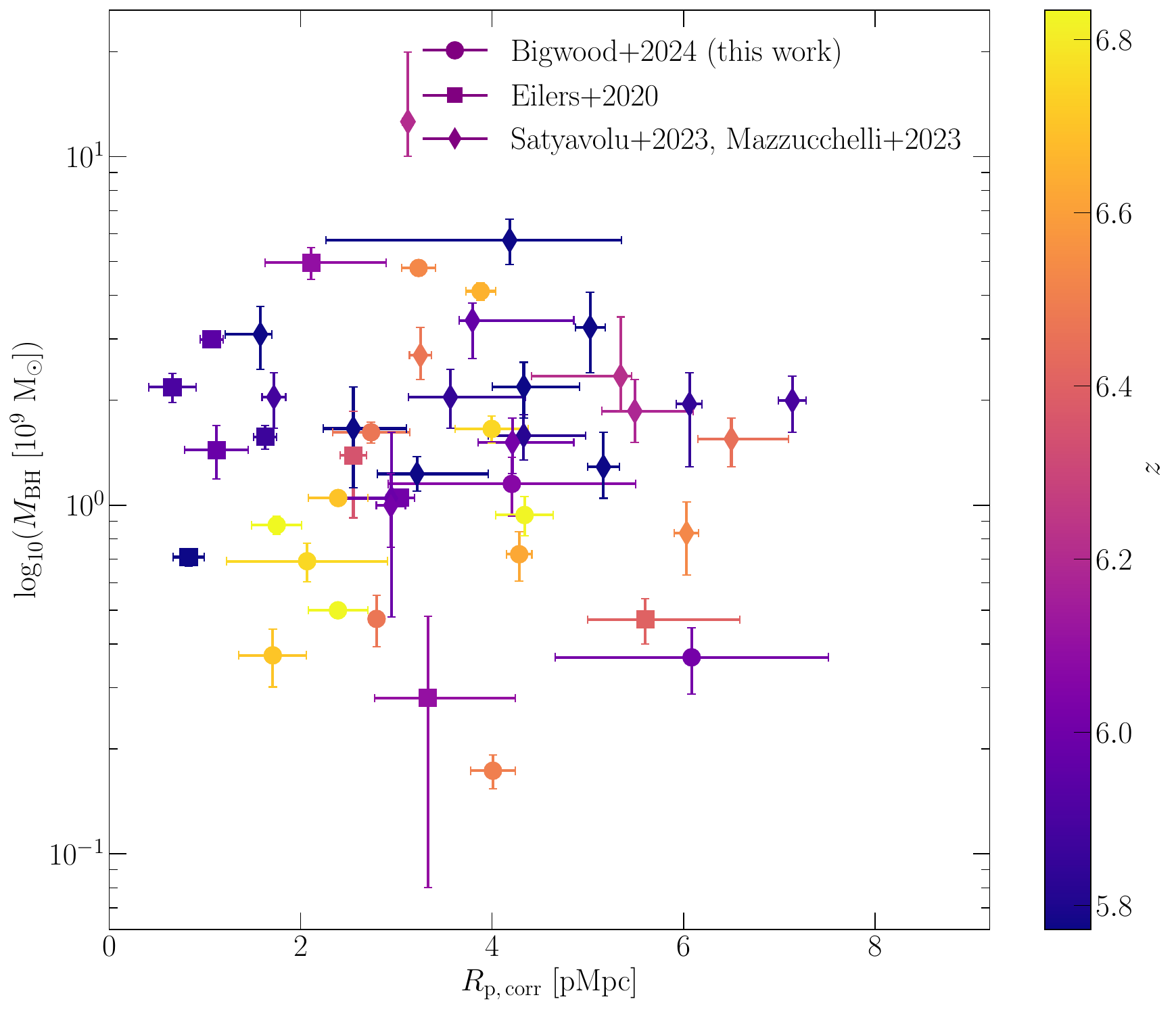}
    \caption{Distribution of quasars in the plane of black hole mass and luminosity-corrected proximity zone size.  The FIRE quasar sample of this work are shown by the circle markers.  We also plot the measurements of \citet{Eilers_2020} (square markers) and \citet{Satyavolu2023} and \citet{Mazzucchelli2023} (diamond markers). 
    }
    \label{fig:massvsproxim}
\end{figure}

\section{Summary}
\label{sec:summary}
In this work, we present a new high redshift quasar sample obtained with FIRE on the 6.5 Magellan/Baade telescope.  All spectra, observed over more than 83 hours, 
are reduced homogeneously using the pipeline \texttt{PypeIt} version 1.7.1.  We estimate the black hole masses of this quasar set, and demonstrate that each hosts a billion solar mass SMBH, as expected from previous high-$z$ quasar studies.  We also measure the proximity zone sizes of this quasar sample, since they are known to correlate with the UV luminous quasar lifetimes. We do not find a correlation between the luminosity corrected proximity zone sizes and black hole masses, which suggests that quasar activity is episodic and intermittent. 

Future work will include detailed modeling of the quasars' proximity zones, including the damping wing signature.  This will allow robust estimates of the quasars' lifetimes to be derived, in addition to the neutral fraction of the IGM.  Obtaining lifetime measurements will enable us to study the dependency with black hole mass more robustly.  We note however that given the lack of correlation between proximity zone size and black hole mass found in this work (Figure~\ref{fig:massvsproxim}), we do not expect to find quasar lifetimes correlated with black hole masses in future studies. 

\section*{Acknowledgements}
The authors would like to thank the support astronomers and telescope operators on Magellan for their help and support with the observations for this work. 
Furthermore, we thank Frederick Davies for sharing his PCA components for the quasar continuum modeling. 

This paper includes data gathered with the 6.5 meter Magellan Telescopes located at Las Campanas Observatory, Chile. 

Leah Bigwood acknowledges funds from the The John (Jack) Simpson Greenwell Memorial Fund issued by the College of St Hild and St Bede, Durham University.  

\section*{Data Availability}
The reduced spectroscopic data is publicly available in \citet{Durovcikova2024}. The data reduction pipeline \texttt{PypeIt} is open source. 




\bibliographystyle{mnras}
\bibliography{FIRE} 




\appendix

 \section{Quality of the Continuum Reconstruction}

\begin{figure*}

    \centering
	\includegraphics[width=1.9\columnwidth]{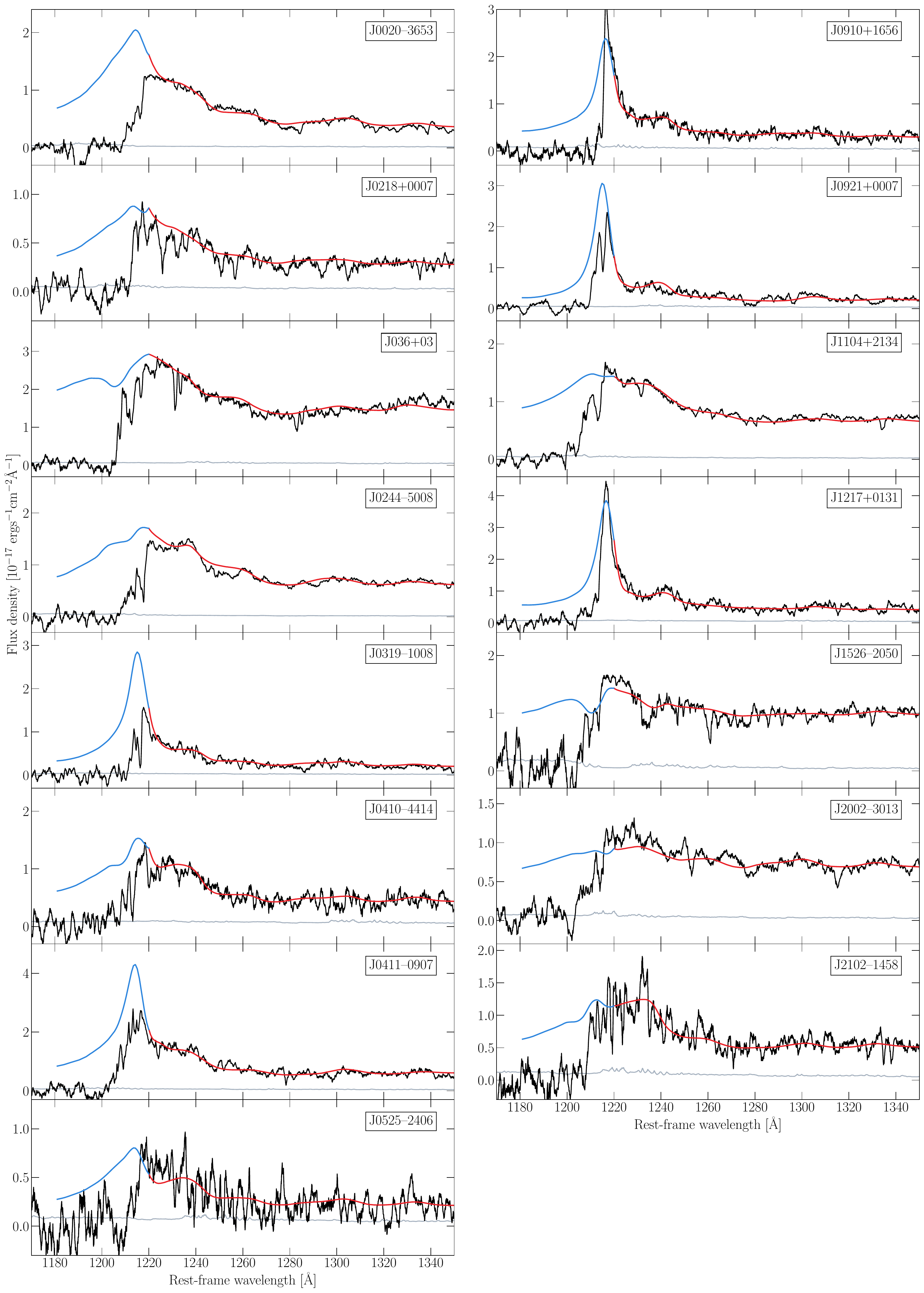}
    \caption{Same as Figure~\ref{fig:continua}, but enlarging the region showing the continuity between the red and blue continuum reconstruction components, in order to better demonstrate the quality of the reconstruction.  We only show the objects for which we utilise the continuum reconstruction in the proximity zone analysis. 
    }
    \label{fig:continua_zoom}
\end{figure*}

In order to better demonstrate the continuity of the red and blue continuum reconstruction components, and thus the quality of the continuum reconstruction, Figure~\ref{fig:continua_zoom} enlarges the region $1170<\lambda_{\text{rest}}<1350$~\AA ~of Figure~\ref{fig:continua} for visual aid.

\bsp	
\label{lastpage}
\end{document}